\documentclass[conference]{IEEEtran}
\IEEEoverridecommandlockouts
\usepackage{cite}
\usepackage{amsmath,amssymb,amsfonts}
\usepackage{algorithmic}
\usepackage{graphicx}
\usepackage{textcomp}
\usepackage{xcolor}
\usepackage{lscape}
\usepackage{multirow, multicol, stfloats}

\usepackage{longtable}
\def\BibTeX{{\rm B\kern-.05em{\sc i\kern-.025em b}\kern-.08em
   T\kern-.1667em\lower.7ex\hbox{E}\kern-.125emX}}

\begin{document}

\title{Distributed architecture for resource description and discovery in the IoT}

\author{\IEEEauthorblockN{1\textsuperscript{st} Meriem Achir }
\IEEEauthorblockA{\textit{LSI laboratory  }
\textit{USTHB University }\\
Algiers, Algeria \\
machir@usthb.dz}
\and
\IEEEauthorblockN{2\textsuperscript{st} Abdelkrim Abdelli}
\IEEEauthorblockA{\textit{LSI Laboratory }
\textit{USTHB University }\\
Algiers, Algeria \\
Abdelli@lsi-usthb.dz}
\and
\IEEEauthorblockN{3\textsuperscript{nd} Lynda Mokdad}
\IEEEauthorblockA{\textit{LACL Laboratory }
\textit{UPEC University }\\
Paris, France \\
lynda.mokdad@u-pec.fr}
}

\maketitle
%Abstract part 

\begin{abstract}
Nowadays, the Internet of Things (IoT) creates a vast ecosystem of intelligent objects interconnected via the Internet, allowing them to exchange information  and to interact. This paradigm has been extended to a new concept, called the Web of Things (WoT), considering that every physical object can be accessed and controlled using Web-based languages and protocols, such as: the CoAP protocol which is becoming the most accepted and suitable protocol in this context. Moreover, the architectures currently proposed for the creation of IoT environments lack efficient and standard support for the discovery, selection and composition of IoT services and their integration in a scalable and interoperable way. To overcome this, in this work, we propose a hybrid and distributed CoAP-based architecture, considering all these aspects by combining the Fog Computing paradigm and structured P2P overlay networks. Furthermore, we describe the different components of our architecture and explain the interaction between them. %Afterwards, we introduce a semantic model, explaining the necessary concepts related to the description of IoT objects and their services in order to enrich it.
\end{abstract}


%Inroduction part
\section{Introduction}

The IoT is a revolutionary technology that is gaining popularity by leaps and bounds. It represents the connection and cooperation of the physical and the virtual worlds through the constitution of a network of objects connected to the Internet capable of collecting and exchanging data and sending them over the Internet, through the use of low-cost computer processing, the Cloud, Big Data, network connectivity, analytics and mobile technologies \cite{Bou}.
These objects can be simple devices, as well as highly complex industrial equipment, such as: sensors, smartphones, home appliances, cars, thermostats, baby monitors, etc. With more than 10 billions IoT devices connected today, experts expect this number to rise to 22 billions by 2025. In this hyper-connected world, digital systems can record, monitor and adjust every interaction between connected objects. 
These physical entities provide IoT with services that can be used to accomplish many different tasks. However, they are limited in terms of resources such as: storage capacity, processing and computation, battery power, etc.
 To ease the interactions in the IoT domain, the concept of IoT gateways has been introduced as an intermediary entity between users of IoT applications and constrained devices, enabling the registration, the discovery and the access to IoT resources. 
 Such a gateway can be viewed as a broker that exposes different resources and handles application requests while taking into account the possibility that IoT objects may not be available. 
 Many approaches for the deployment of these gateways have been proposed in the literature \cite{achir2022}, which consist mainly in placing the gateways in the Cloud to ease the deployment and to ensure scalability, flexibility,  availability of resources (e.g. storage, CPU, etc.) for data analysis and low-cost management \cite{mahmood2018}. 
 However, as the physical distance between data centres and users increases, so does the transmission latency and hence the response time \cite{mahmood2018}. Therefore, the Fog Computing paradigm  has emerged to extend the traditional Cloud Computing and cover its limitations. This concept aims at bringing intelligence, processing and data storage closer to the edge of the network in order to provide IT services faster and closer to the interconnected smart objects that are part of the IoT.
    Such an architecture offers several benefits, including: reduced latency and network bandwidth by improving quality of delivery and enhancing operational efficiency as well as faster real-time processing while providing better support for mobile communications. The placement of IoT gateways in Fog Computing near to IoT devices is already being explored and recognized as a major application scenario for this paradigm \cite{tanganelli2017}.

Furthermore, IoT objects are often incompatible with standard Internet protocols, such as HTTP, which require resources-rich equipment. Among the proposed solutions to address this issue, the Constrained Application Protocol (CoAP), which is emerging as a widespread standard. However, the major limitation of CoAP is that it allows only simplistic static registration and syntactic queries and provides only keyword-based string matching (exact matches), which is not common in the real world scenarios, and appears to be highly inadequate to provide automatic and intelligent resource discovery in IoT. 
The integration of semantic aspects within CoAP can overcome this limitation. Several works have been proposed in this sense by exploiting the use of reference ontologies to enrich the attributes of the CoRE Link Format by annotating objects, and services. However, the majority of these solutions focus only on sensors using the SSN ontology and neglect other types of objects \cite{Gramegna2013}\cite{Ruta2017}. Also, the proposed semantic models do not present all aspects related to objects and their services, such as: QoS, context attributes, location, etc. Furthermore, the process of mapping concepts from the semantic model to CoAP is not clearly defined and is at a preliminary stage. 

Moreover, implementing an efficient discovery system needs first to design its architecture, by describing the structure and the behavior of each of its modules in order to meet all the technical and operational requirements of the system. 
This architecture shows how a system is decomposed, how the processes interact, or the different ways in which system components are distributed across the network.  

In this work, we are interested in CoAP-based architectures \cite{cirani2014}\cite{tanganelli2017}\cite{Gramegna2013}\cite{Ruta2017}.  Within this context, all the existing solutions defined in the literature focus only on the resource discovery and access process, besides the lack of several modules, including: composition, selection, indexing, etc.
To this aim, we propose a hybrid distributed CoAP based architecture for the discovery, the selection and the composition of IoT services using web technologies in a constrained IoT environment. In the proposed solution, we combine the Fog Computing paradigm and structured P2P overlay networks to provide a judicious placement for the Fog nodes (IoT gateways). We describe each component of this architecture and explain the interaction between them.

% Also, we expose the architecture of the framework hosted by each gateway. %Moreover, we introduce a semantic model to explain the necessary concepts related to the description of IoT objects and their services. 

% {  \color{red} Est ce que cette architecture est nouvelle par rapport à la literature ? Si oui qu 'est ce qui la différencie de l'existant ? et quels seraient ses atouts dans le contexte où elle pourrait etre appliquée ?
 
% Autre chose est ce que vous introduisez  la structure d'INdex ds cette architecture ?}

The remainder of this article is organized as follows. Relevant articles are presented and discussed in Section 2.  Section 3 is devoted to present our proposed architecture and explain its components in detail. %Section 4 details the proposed semantic model to represent things and their services including context and QoS attributes. 
The last section concludes this paper and highlights the future research directions. 

\section{Related works}
The IoT environment is composed of a large number of objects connected to each other, providing a variety of heterogeneous services. For this purpose, many techniques have been defined to describe and discover these objects and their services, while  several protocols (e.g.: DNS-SD, MQTT, CoAP, REST, etc.), have been designed to meet the requirements of resource-constrained devices and allow them to communicate with each other and to automatically discover services without human intervention. 
Among the solutions based on these protocols, we can cite: the work presented in \cite{klauck2012}, where authors introduced a DNS-SD/mDNS-based SD technique which operates on Contiki OS embedded devices and is initiated by sending a record containing the searched service name. In the same regard, Stolikj \emph{et al.} have extended the mDNS/DNS-SD protocol with a context-based model for describing and discovering services \cite{stolikj2016}. 

Moreover, Pareira \emph{et al.} developed a distributed MQTT-based resource discovery architecture for M2M communications, providing plug and play capabilities for network devices to enable zero-configuration networking \cite{pareira2019}. Similarly, Venanzi \emph{et al.} proposed an MQTT-based node discovery protocol, considering MQTT brokers as fog nodes to trigger turning on/off the BLE interfaces of the surrounding objects (battery-powered IoT nodes) \cite{venanzi2018}.

The choice of the protocol to consider depends strongly on  the application goal. In the context of the WoT, every physical object can be accessed and controlled using Web-based protocols, including: the CoAP protocol which is emerging as a widely used standard for many reasons. One of the main advantages of this protocol is that it is REST-based and therefore allows interoperability with HTTP and the RESTful Web Services through simple proxies \cite{achir2022}.
Thus, it can easily be integrated within the web. It shares the same methods as HTTP (e.g., GET, POST, PUT, DELETE, etc.) and uses URIs to invoke services. In addition, this protocol is designed for constrained environments and has a very light and simple packet structure with binary data representation. It thereby enjoys a lower overhead as well as a reduced bandwidth requirements and parsing complexity. Moreover, CoAP supports both request/response and resource/observe models and allows the definition of context and QoS attributes as key-value pairs . 

In this work, we are mainly interested in architectures based on this protocol. Among the CoAP-based solutions proposed in the literature, we can cite the work presented by Ferdousi \textit{et al.} in \cite{ferdousi2019}, where they proposed a Cloud-based middleware, called LOAMY, following a hybrid model which combines two forms of the CoAP protocol (centralized and distributed) for service discovery. The client request is sent to the Refinery Node to analyze its type. If the directory node address is known, the request is forwarded to the Border Router in the distributed form, otherwise it will be transferred to the Client Request Handler (CRH) in the centralized form.  The CRH of the context responds to the request either with an acknowledgement, if the service providers or services are not available, or with a payload containing the requested service, if available. 

In another work, Cirani \emph{et al.} introduced a P2P-based architecture for self-configurable and scalable SD at local and global scales. This architecture relies on the use of IoT gateways which interact with other IoT nodes through CoAP and may act as both CoAP client and CoAP server. These gateways are used to store and retrieve information about IoT nodes and their provided services. They can be federated in a P2P overlay  to provide a distributed and global discovery.  In this solution, authors consider two P2P overlays: the Distributed Location Service to retrieve all the information needed to access a resource identified by a URI, and  the Distributed Geographic Table to retrieve a list of resources matching the geographic criteria \cite{cirani2014}. In the same regard, Tanganelli \emph{et al.} designed an edge-centric distributed architecture to federate different IoT gateways deployed in fog nodes in order to discover and access IoT services \cite{tanganelli2017}. The proposed architecture is based on a structured P2P overlay and thus on DHT (Distributed Hash Table), where information about the IoT resources is stored for global lookup across multiple domains. All the IoT devices in the same domain are managed by the same gateway. The latter is responsible for the registration, discovery and access to IoT resources using the CoRE Resource Directory and CoAP protocol. Moreover, for the implementation of the DHT, the eXtendible Metadata Hash Table (XMHT) has been adopted to exploit its extensible interface that resembles the CoRE Link Format.

Moreover, various research works have been proposed recently to improve interoperability between heterogeneous IoT devices using semantic web technologies and CoAP protocol, we can quote the work presented in  \cite{vandana2020}, where a CoAP-based framework is proposed. The latter is divided  on four layers. The first one is the \textit{Sensor Abstraction Layer} which represents the real-world physical objects exposed through the use of standard interfaces. Secondly, the \textit{Data Access Layer} which comprises three modules: repository, triple store and reasoner. In this layer, the resource descriptions are stored in the repository, while the triple store is used to store RDF. The role of the reasoner is to  match the desired requirement with the most suitable and available services by inferring logical consequences from a set of axioms based on QoS, location and application entity. The third one is \textit{Resource Registration and Discovery Layer} where the CoAP CoRE Link Format is enriched with semantic tags provided by the semantic matchmaker module and the request is resolved by fetching from the repository using SPARQL. Finally, the \textit{Social-ambient Overlay layer} to perform semantic-based group composition by the cluster head. Each resource semantic must have an attribute called Social Relation Type which represents the relationship between the different resources.  
In addition, Djamaa \emph{et al.} have proposed a CoAP-based framework structured around five main components namely: Resource Directory, Device, Border Router, User Interface, and Reference Ontologies Server \cite{djemaa2017}. 

Moreover, Gramegna \emph{et al.} \cite{Gramegna2013} proposed a Semantic Sensor Network (SSN)-based framework that implements a backward-compatible extension of CoAP to support a logic-based matchmaking of semantic-enriched sensors as well as to detect and annotate high-level events from raw data collected from sensing devices using a simple data mining component.  
Similarly, in \cite{Ruta2017}, the authors proposed a semantic Web of things framework, enabling the collaborative discovery of sensors and actuators in pervasive contexts. It is based on a backward-compatible extension of CoAP, supporting advanced semantic matchmaking via non-standard inference services and event annotations using a simple data mining method.
 
 Finally, Jin \emph{et al.} designed and implemented a Semantically Enhanced CoAP Gateway (SECoG), located between Low-Power Lossy Networks and the Internet, which supports the simple and complex CoAP service mashups via a single HTTP request \cite{Jin2019}. Also, two HTTP request templates (one for simple mashups and the other one for complex mashups) were presented to allow users to specify the description of a desired service mashup. In this solution, a group generating method was proposed to group services which are similar to the required ones and a semantic similarity calculation method was designed to provide partial correspondence results in case where there are no exactly matching servers.

\section{The proposed architecture}

\begin{figure}%[ht]
\centering
\includegraphics [scale=0.4]
%[width=1.1\textwidth]
{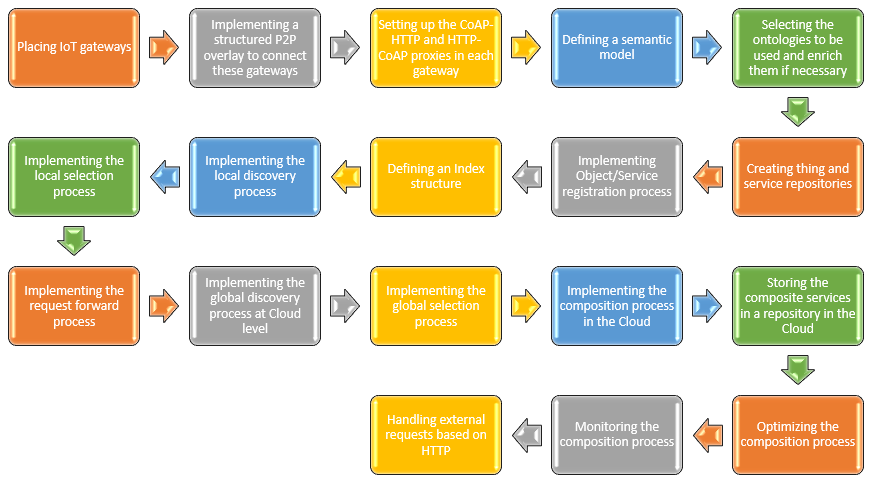}
\caption{The roadmap of the proposed architecture.} 
\label{roadmap}

\end{figure}

In this section, we describe the proposed architecture for the discovery, selection and composition of services, as well as its components and the interactions between them. A roadmap of the creation of this solution is presented in Fig. \ref{roadmap}. First of all, the IoT objects and the network infrastructure must be installed and the gateways, representing the Fog nodes, must be placed and organized as a structured P2P overlay. Then, CoAP-HTTP and HTTP-CoAP proxies have to be set up at each gateway. Besides, a semantic model needs to be defined by considering appropriate basic ontologies to be used.  The latter can be enriched, if necessary, to annotate and improve the description of the IoT objects/services  that are stored in their respective repositories.  For the local discovery process, an indexing structure is defined and a selection process is implemented in order to choose the best services from the discovered ones.

%\begin{landscape}
%\end{landscape}

We have opted for a distributed and modular architecture to bring more flexibility and resilience into IoT environments and to distribute the workload on a number of interconnected cooperating nodes. This type of architecture has several benefits, such as: resource sharing, fault-tolerant, scalability, flexibility and robustness, etc \cite{achir2022}. As depicted in Fig.\ref{distArchi}, our architecture is mainly composed of three layers, which are:

\subsection{IoT Layer}
 This layer includes physical IoT devices equipped with sensors, actuators, RFID tags, etc. and capable to communicate. These objects publish and share their data and services in the nearest Fog nodes where they will be processed. They can be programmed to send their descriptions and those of their services, once connected to the gateway to which they are linked, via the POST method of the CoAP protocol.
   % {\color{red}  est ce qu il y a des technologies specifiques à utiliser au niveau objet : Agents, Protocole ou autre mecanisme ? pour assurer la communication et la découverte et l execution des services }

\subsection{Fog Layer }
\begin{landscape}
\begin{figure}%[ht]
\centering
\includegraphics %[scale=0.47]
[width=1.1\textwidth]
{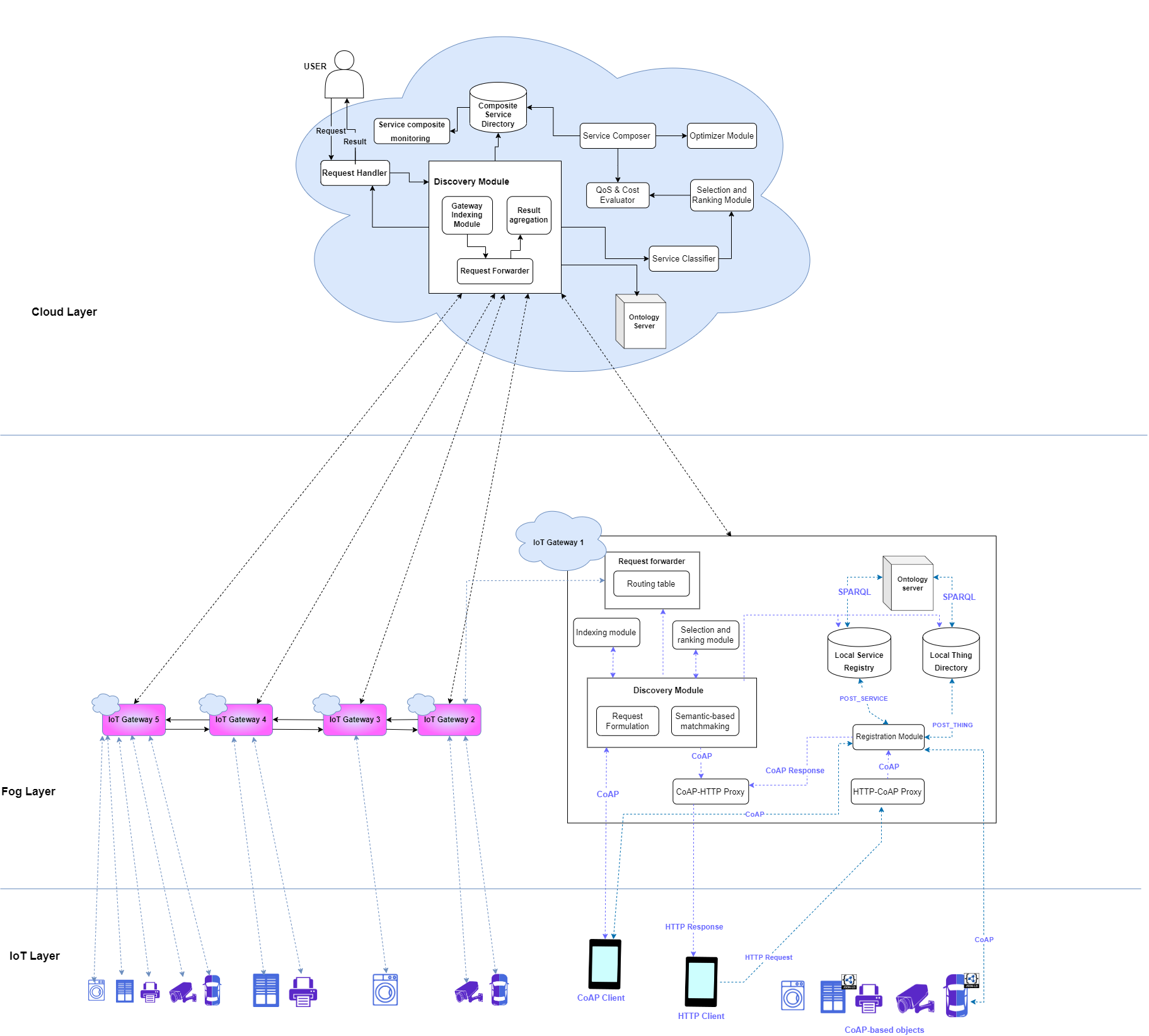}
\caption{The proposed distributed architecture for IoT Resource discovery, selection and composition.} 
\label{distArchi}

%{\color{red} %pourquoi l utilisateur doit passer par le cloud forcemment pourquoi pas par le fog ? Sinon pourquoi les fog nodes s'echangent ils des requetes puisque ds votre schema c'est transversal.

%On ne voit pas le coté internet HTTP et COAP
%Si vous donnez une architecture et que les modules ainsi que leurs agents ne sont pas représentés ça n a aucune utilité}

\end{figure}
\end{landscape}

The Fog layer consists of multiple nodes (IoT gateways), deployed over a wide geographical area close to users to bring intelligence, processing and data storage closer to the edge of the network in order to deliver services faster and closer to the interconnected intelligent objects that are part of the IoT. IoT gateways manage these objects and their services, control their availability and enable the local discovery. Also, they have the right to process data and make decisions and this makes the process faster.
In addition, to address latency-sensitive applications by minimizing delays and the number of network hops, a judicious placement of Fog nodes is required. In our solution, we have opted for the use of a structured P2P network based on DHT to link the different IoT gateways together. This choice is motivated by several advantages, including: the support of scalability, reliability and mobility, robustness and efficiency, high availability, no central control and so on. In this model, Fog nodes are considered as peers identified by specific known logical identifiers. They do not have particular roles. Therefore, they can be both service requesters and providers, so they are equal with no primary administration device at the center of the network. In this layer, at each IoT gateway, a framework is implemented. As shown in Fig \ref{distArchi}, the architecture of this framework is modular, which greatly facilitates its maintenance in case of failure. In the following, we describe each module of this architecture. 

\begin{itemize}

    \item \textbf{Local Thing Directory (LTD): }It is a directory containing the description of IoT devices and their metadata, following the JSON-LD format. It offers an API to create (POST\_THING), read (GET\_THING), update (PUT\_THING) or delete (DELETE\_THING) these objects.
    \item \textbf{Local Service Registry (LSR): }It is a directory of registrations describing resources hosted on IoT devices, offering several simple services in JSON-LD format. It implements a set of methods, allowing devices to register their services (POST\_SERVICE), maintain them (PUT\_SERVICE and DELETE\_SERVICE) and retrieve them (GET\_SERVICE). An SR can be logically segmented into sectors (clusters) to facilitate the search for the appropriate service.
    \item \textbf{Ontology server: }It is a web server containing all the reference ontologies used in our work and allows access to them (lightweight version). The information of these ontologies can be retrieved or updated via the SPARQL language.

    \item \textbf{HTTP-CoAP proxy: }It allows HTTP clients to access resources on CoAP-based servers via an intermediary.

    \item \textbf{CoAP-HTTP proxy: }It enables  the translation and the mapping of  CoAP requests into  HTTP ones
    \item \textbf{Registration module: }It is responsible for registering objects and their services in the two directories via dedicated APIs. As aforementioned, an object can send its description and those of its services to the selected gateway. Similarly, a user can do this following the following steps: 
    \begin{itemize}
        \item First, we distinguish two kinds of users: CoAP users and HTTP users.
        \item If an HTTP user sends a registration request (HTTP POST method), he must pass through the \textit{HTTP-CoAP proxy} which then forwards it to the \textit{Registration Module}. Otherwise, if it’s a CoAP user, the communication is direct and no translation is required.

        \item Then, this module extracts the object description from the registration request to invoke the POST\_THING method to register the description into the LTD. Once the object is saved, its ID is returned to the sender (i.e. \textit{Registration module}).
    
        \item Once the object ID is retrieved, this module invokes the POST\_SERVICE method to register one by one the services offered by this object in the LSR (it also saves the object ID in the service description).
    
    \end{itemize}
    
    \item \textbf{Discovery module: }It is responsible for the search and the discovery of objects/services according to the functional capabilities matching the properties which are specified by the user in his discovery request.
    
    \begin{itemize}
        \item \textbf{Request formulation: }This component extracts the necessary information from the CoAP request sent by the user and formulates a SPARQL query that will be executed by invoking the methods offered by the LTD and the LSR.
        
        \item \textbf{Semantic-based matchmaking:  }This component calculates the semantic correspondence (semantic similarity) between the information provided by the user and the descriptions of the objects/services existing in the local directories.
    \end{itemize}
    
     A list of descriptions, meeting the user's requirements, is retrieved and subsequently sent to the \textit{Selection and ranking module}. When a service is not found on a given gateway, it is forwarded to the \textit{Request forwarder} module. 
 
    \item \textbf{Indexing module: }This component is used by the \textit{Discovery Module} to accelerate and optimize the query processing.   
     
    \item \textbf{Selection and ranking module: }This module receives the list of the candidate services and selects the most appropriate service(s) based on the non-functional properties, which are known as the QoS parameters, e.g. price, security, reliability, delay, etc.
 
    \item \textbf{Request forwarder: }This component is responsible for routing the user request to another node (IoT gateway), if no matching service is available locally, via the \textit{Routing table}. 
 
    Each peer has a local \textbf{Routing table} which stores a list of routing rules and is used by the forwarding algorithm to redirect the query to other nodes of the overlay. It is initialized when a gateway joins the overlay, using a defined procedure. The peers periodically exchange updates of their routing tables as part of the maintenance of the overlay. Also, the join and leave of the IoT gateways are dynamically managed by the P2P overlay without requiring any configuration on a central server.
 
\end{itemize}
    
%{\color{red} c'est un peu flou avc le description que vous donnez plus bas ds le could ? ds la partie cloud vous dites que les requetes viennent du cloud puis sont diffusées à tous les fogs ? et ici vous dites qu un fog node peut generer une requete et l envoyer à ses voisins ?  il faut bien clarifier le contexe de ces deux scenarios, je pense  vous parlez de deux types de requetes differentes ? Bref il faut que expliquer qu'elle type de requete doit passer par le cLOUD et ceux qui sont traitees au niveau de la couche FoG }

Moreover, there are different protocols that build structured P2P networks: Content Addressable Network (CAN), Tapestry, Chord, Pastry, Kademlia, etc., but in this work, we do not specify any type of these protocols since they only use the lookup function that is provided by any type of DHT.

%{\color{red} Quelle technlogie est utilisée pour la description des services? on devrait avoir des annuaires distribuées non ? Vous ne parlez pas du tout de COAP ici ? alors qu il me semble qu il intervient à ce niveau ?}

\subsection{Cloud Layer} 
The service composition process is complex and involves several techniques, such as: discovery, selection, classification, etc. 
In our architecture, this process is performed  at the \emph{Cloud Layer} level for many reasons, including: the availability of unlimited computation ans storage resources, the flexibility and scalability, as well as the adoption of the pay-as-you-go model which can dispense users from maintaining high-end hardware resources. In the proposed solution, if a service is not found at the fog layer, the node forwards the request to the \emph{Discovery module} at the Cloud layer. First, this module searches in the \emph{Composite Service directory} to find a combination that can answer this request if it has already been processed, otherwise the composition process is triggered, by following next steps:

\begin{itemize}
    \item The discovery module selects the gateways to which it will forward the request (simultaneously) via \textit{Gateway indexing module}, containing all the gateways' information as well as their locations, and then it forwards it for processing.
    \item After performing a local search, each gateway returns a list of its available services, matching exactly or partially (i.e., only part of it) this request, to the \emph{Discovery module}. The latter aggregates the several returned results from each gateway and sends the list of candidate services to the \emph{Service Classifier} component.
    \item This module then divides these services into several clusters according to their functional properties.
    
    \item The selection and ranking process is triggered afterwards (\emph{Selection and Ranking Module}), by considering and evaluating the non-functional properties, i.e. the QoS parameters, (\emph{QoS \& Cost Evaluator}) to choose and select the most adequate and performing services within each class of services matching the functional parameters.

    \item The list of selected services is sent to the \emph{Service Composer} module to orchestrate the composition, but before that, the QoS of this combination is evaluated ( \emph{QoS \& Cost Evaluator}) and an optimal execution plan is established based on the set of selected services and the optimization techniques provided by the \emph{Optimizer Module}.

    \item Finally, the composite service will be saved in the \emph{Composite Service directory} to avoid reprocessing it each time. This directory is maintained by the \textit{Service composite monitoring} module, which can replace a simple service by another if it's no longer available. 

\end{itemize}

Furthermore, we take into account external user requests received at the Cloud layer by the \emph{Request Handler} component which will trigger the discovery and selection process as previously described. Besides, at this level the \textit{ontology server} contains complete and enriched ontologies used for more complex query processing, when required.
%{  \color{red} On devrait trouver une structure speciale pour sauvegarder les services webs composés (déja elaborées et delivres)  avec leurs parametre QdS et QdE pour eviter de retraiter certaines requetes

%Est ce qu il y a module d'orchestration ou de choregraphie des services composés ? ou est ce qu il intervient Cloud Fog ?}

\section{Conclusion}
In this work, we have designed a hybrid and distributed architecture for the discovery, the selection and the composition of services offered by IoT objects using web technologies (e.g. CoAP, HTTP, etc.) in a constrained (IoT) environment. This architecture is modular and is divided into three layers: IoT layer, Fog layer and Cloud layer. We described the functionalities of each component of this solution and explained the interaction between them. Future work will lead us to implement the different components to evaluate its performances.

%{\scriptsize

%}

\end{document}